\documentclass[aps,prl,twocolumn,showpacs,groupedaddress,preprintnumbers,amsmath,amssymb]{revtex4}
\usepackage{graphicx}
\usepackage{dcolumn}
\usepackage{bm}
\usepackage{epstopdf}
\usepackage{color} 

\begin{document}

\preprint{03/2012-PRL}

\title{Theory of single-molecule experiments in the overstretching force regime }

\author{Fabio Manca$^1$, Stefano Giordano$^2$, Pier Luca Palla$^{2,3}$, Fabrizio Cleri$^{2,3}$, and Luciano Colombo$^1$}

\affiliation{$^1$Dipartimento di Fisica, Universit\`a di Cagliari, Cittadella Universitaria, 09042 Monserrato (Ca), Italy}
\affiliation{$^2$Institut d'Electronique, Micro\'electronique et Nanotechnologie (CNRS UMR 8520)} \affiliation{$^3$Universit\'e de Lille I, 59652 Villeneuve d'Ascq, France}

\date{\today}

\begin{abstract}
We present a statistical mechanics analysis of the finite-size elasticity of biopolymers, consisting of domains which can exhibit transitions between more than one stable state at large applied force. The constant-force (Gibbs) and constant-displacement (Helmholtz) formulations of single molecule stretching experiments are shown to converge in the thermodynamic limit. Monte Carlo simulations of continuous three dimensional polymers of variable length are carried out, based on this formulation. We demonstrate that the experimental force-extension curves for short and long chain polymers are described by a unique universal model, despite the differences in chemistry and rate-dependence of transition forces.
\end{abstract}


\maketitle
\setlength{\parskip}{0 pt}

Dynamic force spectroscopy by means of the atomic-force microscope (AFM), laser- or magnetic-tweezers apparatus, or the biomembrane force probe, allows the direct probing of the elasticity of individual molecules, and as such has rapidly become a mainstay of biophysical research \cite{ritort,cleri,life,attila}. These mechanical devices are quite different from one another, one prominent difference being their equivalent stiffness, in the range of $10^{-4}-1$ pN/nm for tweezers, vs. $10-10^5$ pN/nm for the AFM \cite{attila}. The typical experiment is a mechanically-induced unfolding of a biological polymer made of $N$ domains, e.g. a polysaccharide such as dextran \cite{rief1}, a protein such as titin \cite{rief2}, a DNA or RNA strand \cite{busta1}, and so on.  As a function of increasing force levels different mechanical response regimes are observed, beginning with the entropic unfolding of the polymer chain (now well understood in terms of simple worm-like chain (WLC) or freely-jointed chain (FJC) models \cite{marko}); to the linear-elastic extension of the straightened chain; to the so-called \emph{overstretching}, typically interpreted as a conformational transformation of the domain geometry; up to the eventual fracturing of the polymer. 

\begin{figure}[b]
\resizebox{9\columnwidth}{!}{\includegraphics{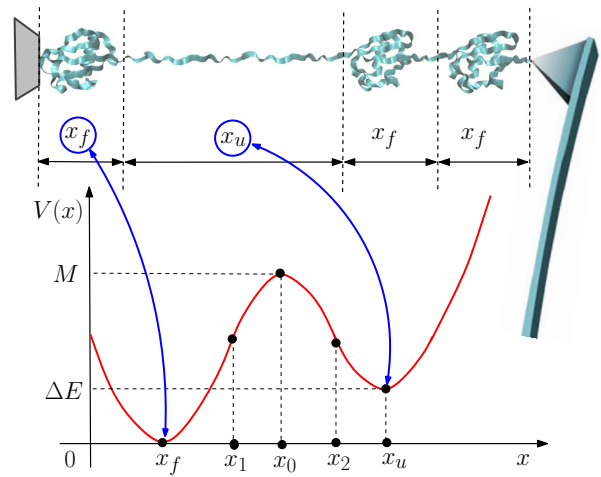}}
\caption{Potential energy function with an energy barrier. Folded and unfolded configurations of the domains are schematically represented. }
\label{conformazioni}       
\end{figure}

In this Letter we provide a robust statistical mechanics foundation to the interpretation of the overstretching regime, which we describe in terms of the internal dynamics of a chain of two--state systems undergoing a conformational transformation, as described by the double-well potential in Fig.\ref{conformazioni}. For the sake of argument we call 'folded' and 'unfolded' the two conformations; however the transformation occurs, more generally, between two principal local minima of the domain free-energy hypersurface (e.g., for DNA it could as well represent the melting transition \cite{cluzel}). Then, we firstly develop a theoretical model describing experiments at constant applied force (a realization of Gibbs ensemble statistics) and we show that the conformational change must occur simultaneously for all the domains at a given threshold force. 

On the other hand, experiments performed at constant--displacement  are a realization of the Helmholtz ensemble of statistical mechanics. 
In our previous work \cite{manca}, we showed  that the outcome of the two types of experiment converge in the thermodynamic limit of infinite chain length, $N \rightarrow \infty$. 
In practice, real experiments always fall inbetween these two ideal extremes. Therefore, here we focus on the intermediate cases  described by  finite values of the $k_c/k$ ratio, $k$ and $k_c$ being the equivalent spring constant (i.e., stiffness) of the domain and of the pulling device, respectively. We demonstrate by means of Monte Carlo simulations that the typical `sawtooth' pattern \cite{rief2}, observed for the unfolding of large protein domains (such as the Ig units in titin), and the `plateau' or kink \cite{busta1,rief1}, observed in the overstretching of DNA  
and polysaccharides (e.g. dextran), have a common origin in the size-dependence of the polymer response to the external force, the plateau shape being attained in the limit of large $N$. On the same grounds, at a fixed number $N$ of domains, the transition from the `plateau' to the `sawtooth' response is recovered for increasing values of $k_c/k$. Notably, such a behavior of the force-extension curves is universal with respect to the specification of any additional parameters, such as chemical, structural or mechanical constants of the domains.



We work out a simple model containing the minimal ingredients fully describing the overall complex behavior of a polymer chain. It consists of an $N$-domain, non-branched chain clamped at one end, able to describe conformational transitions across an energy barrier. The internal state of each domain is described by a potential energy $V(x)$ which exhibits two minima corresponding to the lengths $x=x_f$ (folded conformation) and $x=x_u$ (unfolded conformation), connected via an energy barrier $M$ at $x=x_0$ (see Fig.\ref{conformazioni}). The energy is written as a $\mathcal{C}^2$ piecewise function, constructed by imposing continuity and differentiability at the joining points $ x_1 $ and $ x_2 $:
\begin{eqnarray}
\label{atratti}
V(x)=\left\lbrace
\begin{array}{ll}
\frac{1}{2}k(x-x_f)^{2}& 0<x<x_{1}\\
-\frac{1}{2}k(x-x_{0})^{2}+M& x_{1}<x<x_{2}\\
\frac{1}{2}k(x-x_u)^{2}+\Delta E& x>x_{2}\\
\end{array} \right.   
\end{eqnarray}

For chosen values of the lengths $x_f$ and $x_u$, the domain spring constant $k$, and the energy difference $\Delta E$ between the two conformations, the other parameters are simply given by: $\delta=x_u-x_f$, $x_{0}=(x_u+x_f)/2+{2\Delta E}/(k\delta)$, $M=(k/4)[\delta/2+2{\Delta E}/(k\delta)]^2$, $x_{1}=x_f+\delta/4+{\Delta E}/(k\delta)$ and $x_{2}=x_u-\delta/4+{\Delta E}/(k\delta)$.
Therefore, this model properly gives a barrier with $x_f<x_0<x_u$ only for $ \vert \Delta E\vert\leq k\delta^2/4 $. 

Upon application of a constant force $f$ to the end of the polymer identified by the position vector $\vec r_N=(x_N,y_N,z_N)$ (the other end being fixed in the origin), the statistics of the fluctuating chain is a realization of the Gibbs ensemble \cite{manca}. 
The ensemble partition function is given by $Z_{f}(f, T) = \int \int_{\Gamma_{N}} e^{-{\widetilde{h}}/k_BT}dq^Ndp^N$, with \(\Gamma_{N} = \Re^{6N}\). The augmented Hamiltonian $\widetilde{h}$ includes the classical kinetic energy of the domains with mass $m$, their total potential energy, and a term, $- f z_N$, describing the applied force along the $z$-axis \cite{manca}. In the framework of the present minimal model, the partition function can be explicitly calculated as:
\begin{eqnarray}
\label{mainp}
&& Z_{f}(f, T) = \left( \frac{2\pi m}{\beta}\right) ^{3N/2}\left( \frac{2\pi }{\beta f}\right)^{N} 
 \\ \nonumber
 &&\times \left\{ \Pi \left(\beta k,\beta f,x_f,0,x_{1} \right)\right. 
  + e^{-\beta M} \Pi \left(-\beta k,\beta f,x_{0},x_{1},x_{2} \right)
  \\ \nonumber
 && \left. \,\,\,\,\,\,+ e^{-\beta \Delta E} \Pi \left( \beta k,\beta f,x_u,x_{2},+\infty \right) \right\} ^{N}
\end{eqnarray}

\noindent with $\beta$=$(k_BT)^{-1}$ and
\begin{eqnarray}
\label{pipipi}
\Pi \left(\alpha,\gamma,x_0,a,b \right)=2\int_{a}^{b}x e^{-\frac{\alpha}{2} (x-x_{0})^2} \sinh{(\gamma x)} dx
\end{eqnarray}

The extension $r$ at a given force is obtained from the partition function as $r=k_B T (\partial \log Z_{f}/\partial f)$.  Since the extension is linearly dependent on $N$, the data for chains of different lengths can be scaled to a single curve upon diving by $N$.

\begin{figure}
\resizebox{0.67\columnwidth}{!}{\includegraphics{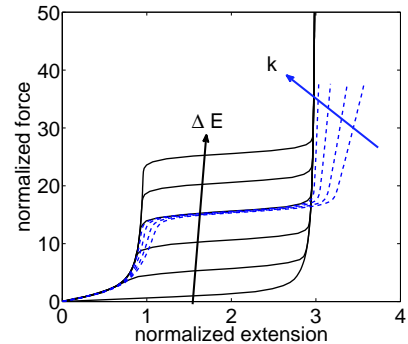}}
\caption{(color online) Force-extension curves for the Gibbs ensemble: normalized force $f/f_\beta$ vs normalized extension $r/(Nx_f)$. The black solid lines correspond to different values of the energy $ \Delta E$=0, 10, 20, 30, 40, 50 $k_BT$ (increasing values from the bottom up) for a fixed spring constant $ k=2000$ $k_BT/\mbox{(nm)}^2$. The blue dashed lines correspond to different values of the spring constant $k$=10, 15, 30, 100 $k_BT/$(nm)$^2$ (increasing values from the right to the left) for a fixed value of the energy barrier $ \Delta E=30$ $ k_BT$. }
\label{conformazioniDE}       
\end{figure}

Figure \ref{conformazioniDE} shows the results of the normalized force-extension curves, $f/f_\beta$ (where $f_\beta^{-1}=\beta x_f$) in terms of $r/(Nx_f)$   
for different values of the energy barrier $\Delta E$=0, 10, 20, 30, 40, 50 $k_BT$ (black solid lines) at a fixed value of $k$=2000 $k_BT/\mbox{(nm)}^2$, and for different values of the spring constant  $k$ (blue dashed lines) at a fixed value of $\Delta E$=30 $k_BT$.
Both sets of curves display a force plateau at $ f \simeq\Delta E/\delta $, for any $ \Delta E>0$, with a normalized width equal to $\delta$. 
In our model, the plateau indicates a transition in the polymer conformation, meaning that for $ f<\Delta E/\delta $ each domain is found in the folded conformation at $x$=$x_f$, while for $f>\Delta E/\delta$ domains are in the unfolded conformation at $x$=$x_u$. i.e., the ensemble of domains respond \emph{cooperatively} to the external force. 
Notably, the value of the plateau force inducing the conformation transition does not depend on the spring constant, $k$, nor on the temperature. Such a result is readily interpreted in the framework of the Bell--Kramers theory \cite{bell}, as the threshold value of force necessary to make the unfolding rate equal to the (reverse) folding one, i.e. lowering the difference $\Delta E$ to zero.
For example, in the case of $ds$DNA, which displays a plateau at $f$=65 pN with a $\delta \approx 2 \AA$, our criterion gives a value $\Delta E$=3.5 $k_BT$, in fair agreement with available experimental data \cite{cluzel, ahsan}. 
A similar plateau was observed for other long chain polymers, such as dextran with $N$=275, $x_f$=0.5 nm, $x_u$=0.56 nm, $\Delta E=13.2$ $k_BT$ \cite{rief1}, for which the simple criterion $f\approx \Delta E/\delta$ gives plateau forces in the range of $\approx$900 pN, as indeed observed \cite{rief3}.

\begin{figure}[ht]
\resizebox{0.67\columnwidth}{!}{\includegraphics{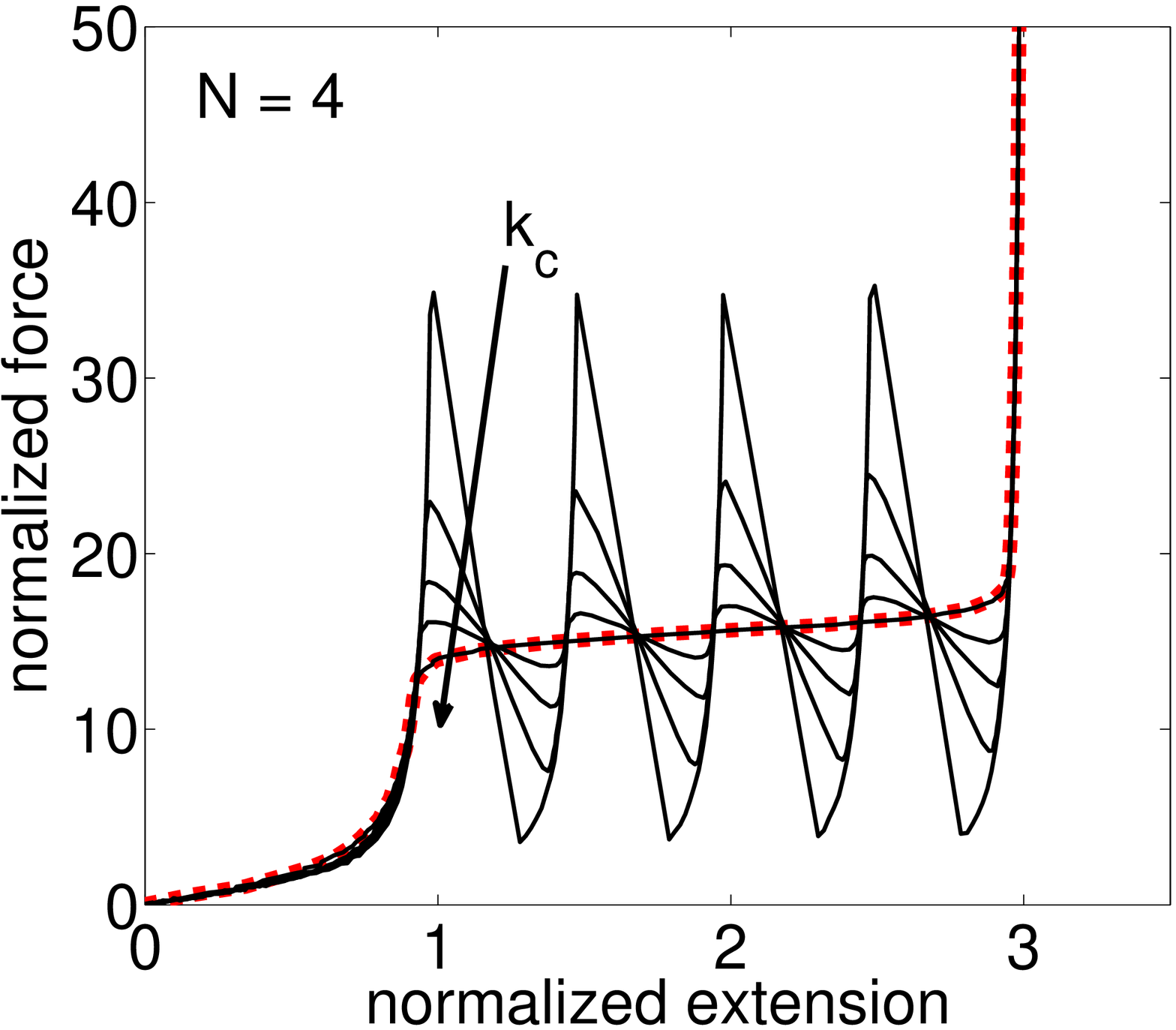}}
\resizebox{0.67\columnwidth}{!}{\includegraphics{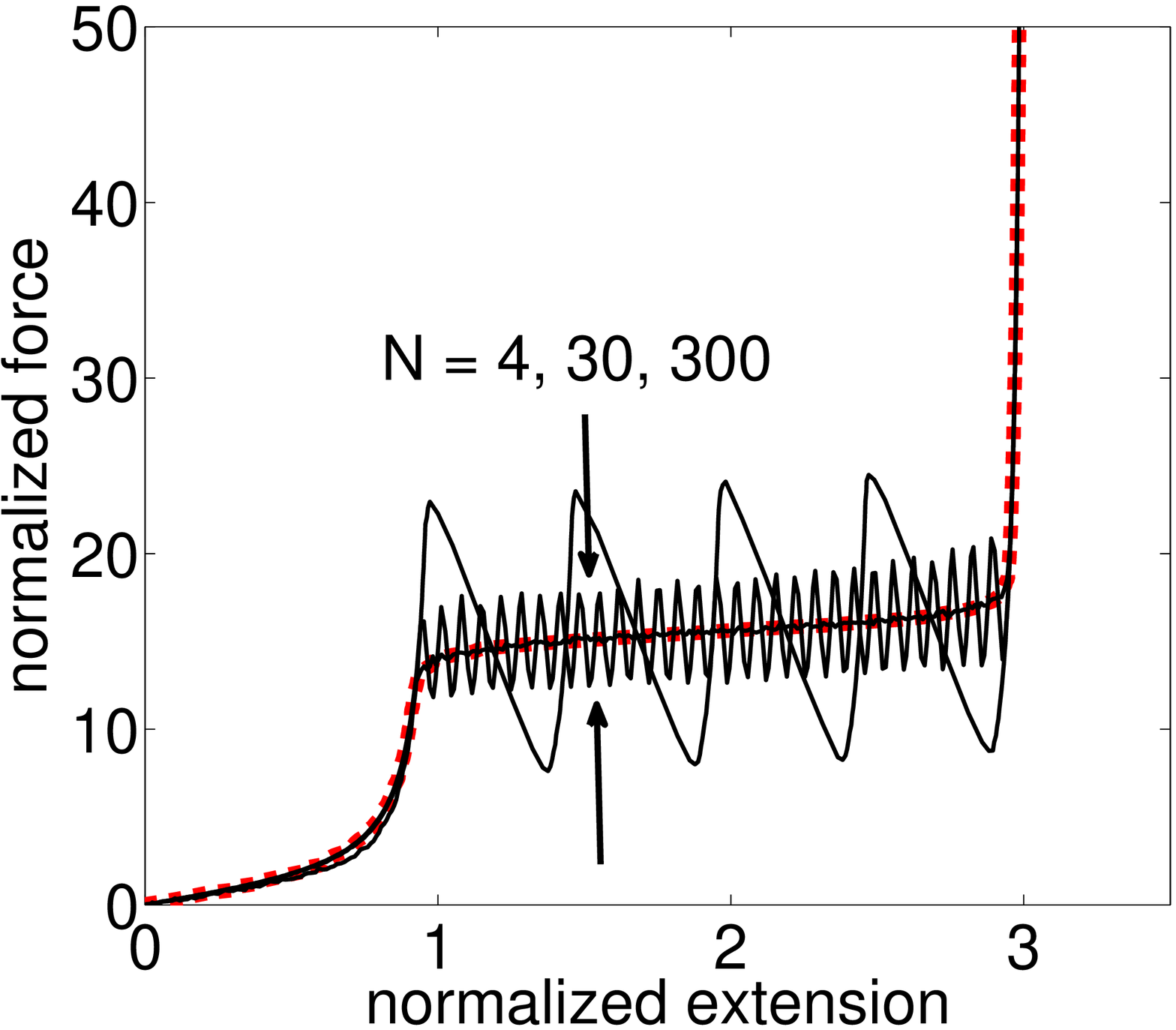}}
\caption{(color online) Monte Carlo force-extension curves at $T$=293 K, for: (top panel) different decreasing values of the device spring constant  $ k_c=$5, 2, 1, 0.5, 0.01 $k_BT/$(nm)$^2$ (from the top down) and $N$=4; (lower panel) increasing number of domains $N=4, 30, 300$ with $ k_c=2$ $k_BT/$(nm)$^2$ (bottom panel). The red dashed line corresponds to the Gibbs ensemble. The remaining parameters are $\Delta E=30$ $ k_BT$, $x_f=2.5$ nm, $x_u=3x_f$ and $k=100$ $ k_BT/\mbox{nm}^2$. }
\label{conf}       
\end{figure}

While Gibbs ensemble statistics are sampled with a constant applied force, a dual situation can be realized by imposing the extension, i.e. by controlling the polymer end-to-end distance. The statistics of the fluctuating polymer in this latter scheme is a realization of the Helmholtz ensemble. As shown in Ref.\cite{manca}, the corresponding partition function $Z_r$ cannot be written in closed form and, as opposed to the Gibbs case, the corresponding extension $r$ is non-linearly dependent on $N$. However, we showed that the partition functions in the two ensembles are formally related via a Laplace transform, and we demonstrated \cite{manca} that they lead to a common force-extension curve in the thermodynamic limit. 

It should be noted that any AFM or tweezers experiment falls in an intermediate regime between the two ideal extremes, of purely constant--force or constant--extension, since either constraint on the terminal domain of the chain is mediated by a mechanical device (such as the AFM cantilever, or the laser--bound microsphere, plus a molecular spacer providing adhesion). The device is characterized by its own effective elastic constant $k_c$, which is coupled in series to the chain of domain springs $k$. In the limit of a soft device, $k_c/k \rightarrow$0, the statistics of the coupled system reduces to the Gibbs ensemble for the isolated molecule fluctuating under a constant force. On the other hand, for a very stiff device, $k_c/k \rightarrow \infty$, one recovers the Helmholtz ensemble for the isolated molecule held at a fixed extension by the fluctuating force \cite{kreuzer}. To describe such a situation, we adopt a Monte Carlo (MC) numerical approach, simulating the stretching of the chain produced by a device with a proper adjustable elastic stiffness. 

In Figure \ref{conf}, top panel, we report the results of the MC simulations at $T$=293 K, for decreasing values of the $k_c/k$ ratio, from 0.05, that is well within the Helmholtz statistics regime, down to $1 \times 10^{-4}$, i.e., approaching Gibbs ensemble statistics. The remaining parameters are set to $N$=4, $\Delta E$=30 $k_BT$, $x_f$=2.5 nm, $x_u$=3$x_f$ and $k$=100 $k_BT/\mbox{nm}^2$, which can be considered representative of a medium--sized, multi--domain chain protein. At large values of $k_c/k$, the domains exhibit a sequence of \emph{independent} conformational transitions to the unfolded configuration, generating a series of $N$ peaks (sawtooth pattern) which closely resemble the experimental results obtained for short chains (e.g., a titin fragment with $N$=8, $x_f$=4 nm, $x_u$=32 nm, $\Delta E$=11.1 $k_BT$  \cite{rief3}). For $k_c/k \rightarrow 0$ the peak-to-valley width, $\Delta f$, of the sawtooth shrinks and the curve approaches the $k_c$=0 cooperative plateau of Gibbs statistics. In substantial agreement with this finding, pulling experiments on native titin by means of optical tweezers \cite{keller}, having a very small equivalent $k_c$ compared to the AFM one, do not reveal the sawtooth pattern, but rather a smooth, monotonic branch reminiscent of the horizontal plateau.

\begin{figure*}
\resizebox{0.67\columnwidth}{!}{\includegraphics{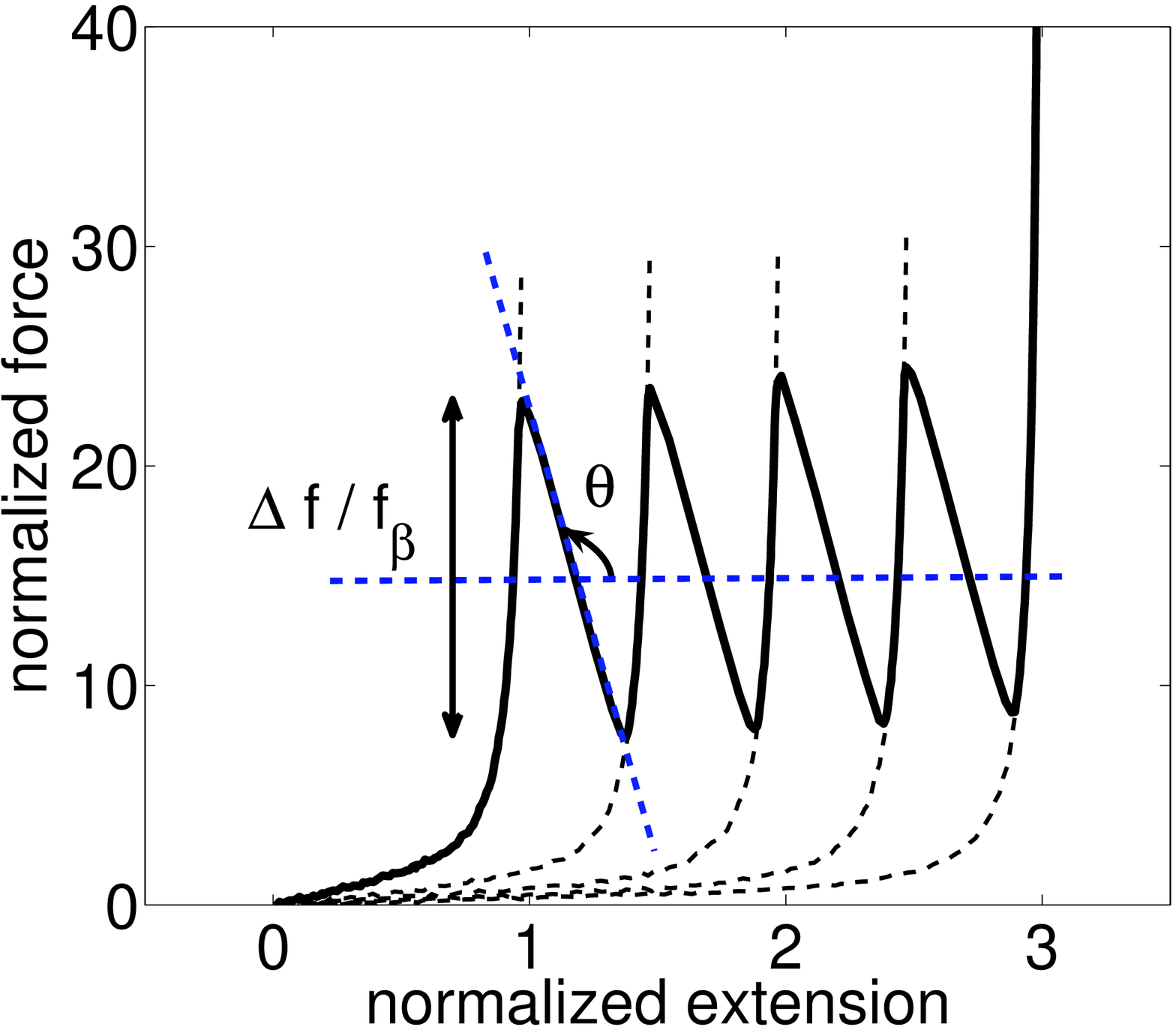}}
\resizebox{0.67\columnwidth}{!}{\includegraphics{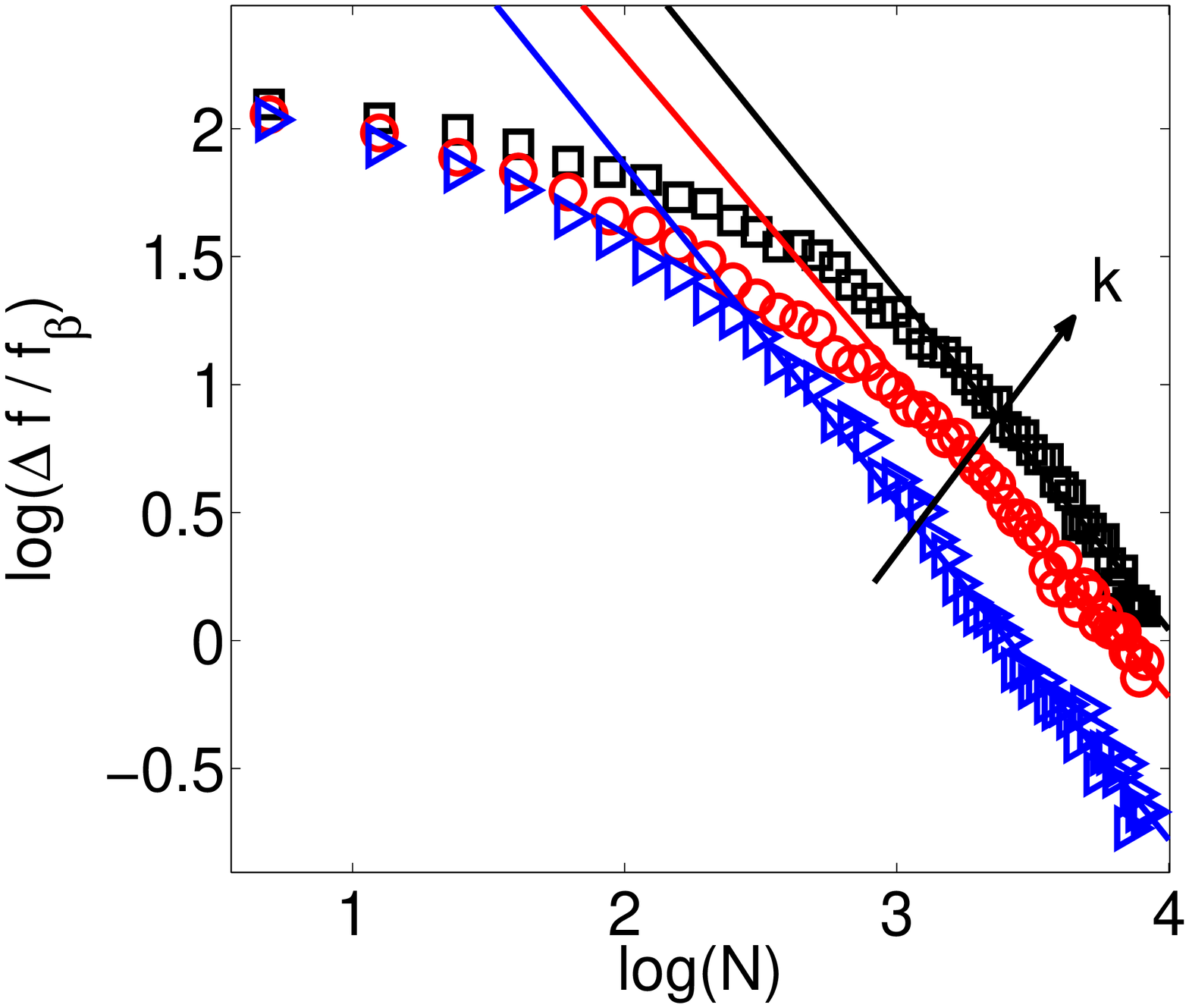}}
\resizebox{0.67\columnwidth}{!}{\includegraphics{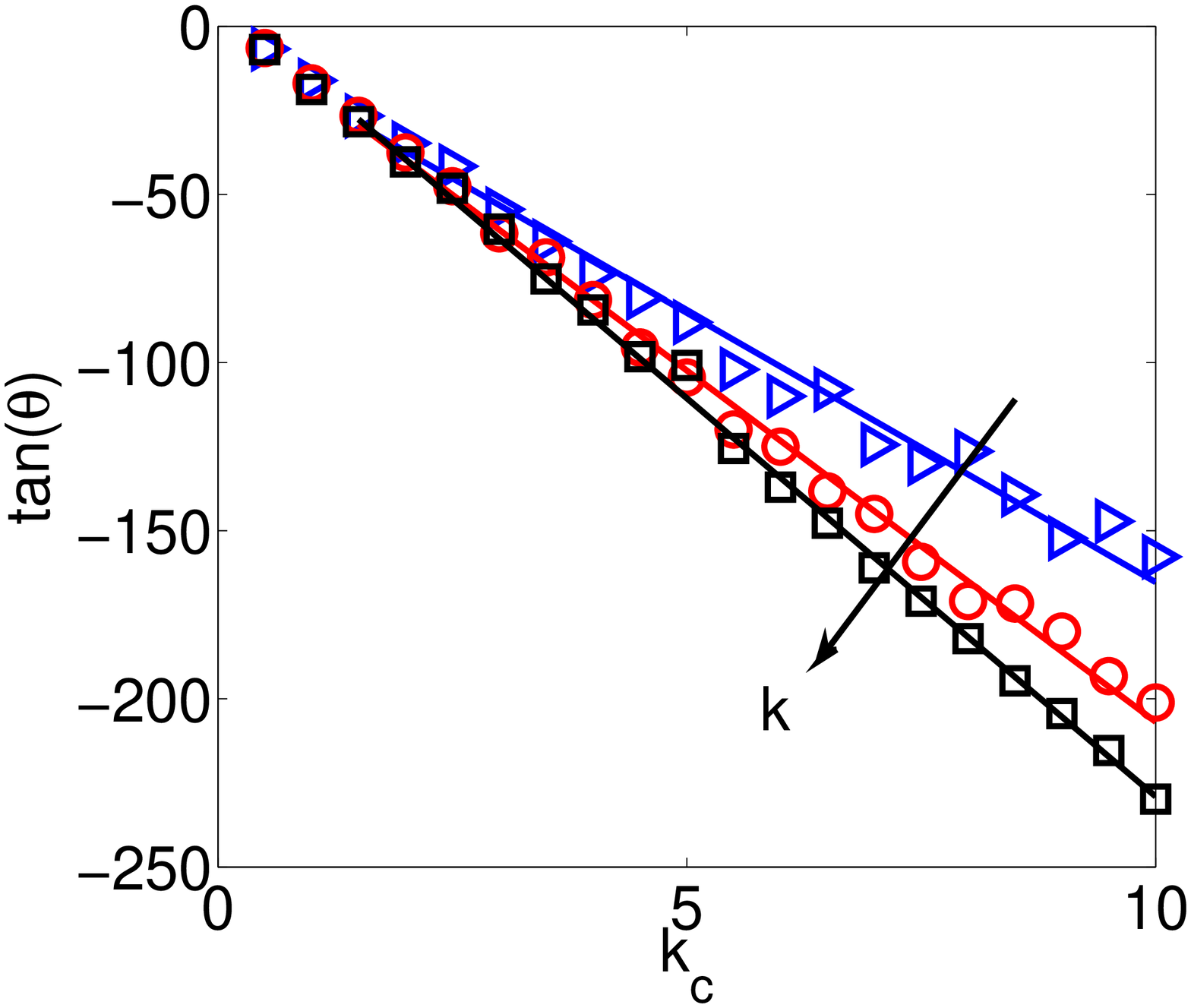}}
\caption{(color online) Left panel: definition of $\Delta f$ and $\theta$ for a typical force-extension curve with $N=4$. Dashed lines for the growing branches fitted to FJC with increasing contour length. Central panel: plot of $\log(\Delta f /f_\beta)$ vs. $\log(N)$ for $k_c$=1 $k_BT/(\mbox{nm})^2$ and $k/k_c$=20, 30, 100 (blue, red and black line, respectively). Right panel: plot of $\tan(\theta)$ vs. $k_c$ for $N$=4, and the same $k$ values. Remaining parameters $\Delta E=30$ $k_BT$, $x_f=2.5$ nm, $x_u=3x_f$ and $T$=293 K.}
\label{tang}       
\end{figure*}

On the other hand, a similar asymptotic trend is observed (Fig. \ref{conf}, bottom panel) when the chain length, i.e. the number of domains, is increased, at a fixed value of $k_c/k$. As $N$ increases, the width $\Delta f$ is decreased until, at a large enough $N$, the force-extension curves approach again the plateau curve of the Gibbs ensemble. It is worth noting that a similar trend was observed in experiments performed on native titin, comprising several hundreds of Ig domains, for which the width $\Delta f$ was of the order of  80 pN \cite{rief2}, compared to the much shorter 8-monomer titin, for which $\Delta f >$ 200 pN. The experiments performed on dextran, a long polysaccharide with $N=275$, \cite{rief1,rief3} whose response to the applied force shows a plateau closer to the typical DNA-like behavior, can also be rationalized on this basis. In summary, we proved that the macroscopically different behavior of small-$N$ polymers (such as titin) vs. long polymers (such as dextran, DNA), as well as experiments done on a same polymer but with devices having widely different stiffness, can be interpreted with the very same unifying model, interpolating between the two extremes of pure Gibbs or Helmholtz statistics.

As observed by several authors, each branch of the sawtooth pattern can be nicely fitted by a sequence of FJC, or WLC curves (see Fig.\ref{tang}, left panel, dashed lines) with a proper value of the persistence length, up to the unfolding of each domain (see e.g. titin \cite{rief2}, spectrin \cite{rief4}, fibronectin \cite{oberh}, synaptotagmin \cite{carrion}). Beyond this point, the force relaxes to a smaller value, until the next curve is met and the force can start rising again upon increasing displacement.


Since the physical origin of the growing branch of the curves is well understood on the basis of FJC or WLC models, we analyzed the decreasing branch, as identified by the common width  $\Delta f$ and angle $\theta$ in Fig.\ref{tang} (left panel) which were extracted from our MC simulations as a function of $N$ and $k_c/k$.

By looking at Fig. \ref{tang} (center), the peak-to-valley width shows a power-law decrease with the chain length, $ \Delta f \sim N^{-\alpha}$, the exponent $\alpha=1.3$ being remarkably independent on the $k_c/k$ ratio. This finding indicates that attainment of the thermodynamic limit is mainly dictated by the thermal force scale, $f_\beta$, and to a much lesser extent by other structural and chemical details of the polymer. It is worth noting that the value of the exponent is in agreement with previous results on mono-stable FJC and WLC models with extensible bonds \cite{manca}.

The last plot on the right of Fig. \ref{tang} reports the behavior of $\tan(\theta)$ as a function of the device stiffness, $k_c$. The observed linear dependence is another remarkable result, completely describing the transition between the two extremes (Gibbs and Helmholtz ensembles), while taking into account all the intermediate cases. For  $k_c/k \rightarrow \infty$ we have $\tan(\theta)\rightarrow -\infty$ or, equivalently, $\theta\rightarrow \pi/2$.
In other words, the decreasing branches of the force-extension curve must be exactly vertical in the case of the Helmholtz ensemble. Notably, both the values of $\Delta f$ and $\theta$ are fully prescribed, i.e. the entire shape of the force-extension curve is uniquely defined, once the free parameters of the model are specified.


In conclusion, we described the statistical mechanics of chain polymers composed by domains with two stable states, subject to a pulling force by a molecular-scale mechanical device. We showed that for short chain length, or large stiffness of the device, the domain response is uncorrelated and originates the typical sawtooth force-extension curve observed in many experiments. On the other hand, upon increasing chain length, or vanishing device stiffness, the response is cooperative and results in the plateau-like curve, also observed in other experiments. Despite the simplicity of the model, such a framework provides a unified picture for such apparently contrasting experimental situations.

\begin{acknowledgments}
F.M. acknowledges the University of Cagliari for the extended visiting grant
and the IEMN, University of Lille I, for the kind hospitality. 
L.C. acknowledges financial support by
RAS under the project 
`Modellizzazione Multiscala della Meccanica
dei Materiali Complessi (M4C)'.
\end{acknowledgments}


\end{document}